# A two dimensional electron gas at the (001) surface of ferromagnetic EuTiO$_3$

R. Di Capua,[1,2,*] M. Verma,[3] M. Radović,[4] N. C. Plumb,[4] J. H. Dil,[4,5] Z. Ristić,[6] E. B. Guedes,[4,5] G. M. De Luca,[1,2] D. Preziosi,[7] Z. Wang,[4] A. P. Weber,[4,5] R. Pentcheva,[3] and M. Salluzzo[2,**]

*[1]Dipartimento di Fisica "Ettore Pancini" Università di Napoli "Federico II", Complesso Monte S. Angelo via Cinthia, I-80126 Napoli, Italy*
*[2]CNR-SPIN Complesso Monte S.Angelo via Cinthia, I-80126 Napoli, Italy*
*[3]Department of Physics and Center for Nanointegration (CENIDE), Universität Duisburg-Essen, Lotharstr. 1, 47057 Duisburg, Germany*
*[4]Photon Science Division, Paul Scherrer Institut, 5232 Villigen PSI, Switzerland*
*[5]Institute of Physics, École Polytechnique Fédérale de Lausanne, 1015 Lausanne, Switzerland*
*[6]Vinča Institute of Nuclear Sciences, Belgrade Serbia*
*[7]Université de Strasbourg, CNRS, IPCMS UMR 7504, 67034 Strasbourg, France*

**Abstract:** Studies on oxide quasi-two dimensional electron gas (q2DEG) have been a playground for the discovery of novel and sometimes unexpected phenomena, like the reported magnetism at the surface and at the interface between LaAlO$_3$ and SrTiO$_3$ non-magnetic materials. However, magnetism in this system is weak and there are evidences of a not intrinsic origin. Here, by using in-situ high-resolution angle resolved photoemission we demonstrate that ferromagnetic EuTiO$_3$, the magnetic counterpart of SrTiO$_3$ in the bulk, hosts a q2DEG at its (001) surface. This is confirmed by density functional theory calculations with Hubbard $U$ terms in the presence of oxygen divacancies in various configurations, all of them leading to a spin-polarized q2DEG related to the ferromagnetic order of Eu-4$f$ magnetic moments. The results suggest EuTiO$_3$(001) as a new material platform for oxide q2DEGs, characterized by broken inversion and time reversal symmetries.

**Keywords**: EuTiO$_3$, ARPES, 2DEG, oxides

The discovery of a surface/interfacial quasi 2-dimensional electron gas (q2DEG) in SrTiO$_3$ [1–3] boosted expectations for novel oxide electronics due to the intriguing and rich physics uncovered, including multiband and possibly unconventional superconductivity [4–6], large spin-orbit coupling (SOC) [7] and large spin to charge conversion efficiency [8]. While magnetic effects were reported [9], several studies attributed them to weak-(para)magnetism induced by extrinsic defects, such as oxygen vacancies [10].

EuTiO$_3$ (ETO) is an antiferromagnetic (AFM) insulator isostructural to SrTiO$_3$ (STO). Like STO, it is characterized by empty Ti-3$d$ bands. In contrast to STO, however, Eu$^{2+}$ magnetic moments order anti-ferromagnetically below 5.5 K and form a spin-polarized band 2 eV below the Fermi level with 4$f$ orbital character. The AFM order reverts into a ferromagnetic (FM) one ($T_c$=6-8 K) by doping or lattice strain [11, 12]. Due to combined effects of SOC and spontaneous Zeeman field, doped-ETO exhibits a bulk band structure with Weyl nodes and a topological Hall effect [13, 14]. These exotic properties may induce intriguing physical phenomena in confined q2DEGs at the ETO surfaces and interfaces.

Here, by employing angular resolved photoemission spectroscopy (ARPES) on in-situ grown thin films, we show that the (001) surface of ferromagnetic ETO hosts a q2DEG. Density functional theory (DFT) calculations suggest a spin-polarized q2DEG related to the FM order of Eu-4$f$ magnetic moments.

We studied ETO films with thickness of 2, 5 and 15 unit cells (uc) deposited *in-situ* by reflection high-energy electron diffraction assisted pulsed laser deposition on TiO$_2$-terminated (001) STO single crystals at the Surface/Interface Spectroscopy (SIS) beamline of the Swiss Light Source (Fig. S1 in Supplementary Information). We also studied a 15 uc ETO sample deposited on a SrRuO$_3$ buffer layer to improve ground contact. After the deposition, the samples were annealed in ultra-high-vacuum (UHV) at 600 °C for one hour and then in-situ transferred to SIS ARPES end-station. The UHV annealing does not perturb the 1x1 surface structure, as confirmed by low energy electron diffraction (Fig. S1 in Supplementary Information). X-ray diffraction shows that the ETO films grow coherently, with almost no strain, and have a c-axis (3.92±0.01 Å), close to that of bulk STO (Fig. S1d-f in Supplementary Information). Finally x-ray magnetic dichroism and SQUID magnetometry data show that the films are ferromagnetic with a $T_c$<10 K and a magnetic moment close to the value of 7 $\mu_B$/Eu expected for $Eu^{2+}$ (Fig. S2 in Supplementary Information).

The ETO surface state was compared to STO-qDEGs with a controlled 2D carrier density $n_{2D}$ created either by mild $Ar^+$ sputtering and low pressure oxygen annealing (STO-A), or just by UHV annealing (STO-B) of $TiO_2$ terminated (001) STO single crystals [15, 16]. To assess the nature of the (001) ETO surface state, in Fig. 1 we report in-plane momentum ($k_x$-$k_y$) Fermi surface (FS) cuts of 2 and 5 uc ETO films deposited on STO (Figs. 1a-b), of a 15 uc ETO film deposited on buffer $SrRuO_3$ (Fig. 1c), and of an STO single crystal (Fig. 1d). Data were acquired at 15 K with photon energy $h\nu$= 85 eV, right-handed circular (C+) polarization and an out-of-plane $k_z$ momentum close to the $\Gamma_{103}$ reciprocal lattice point of the equivalent bulk periodicity [15–17]. FS-maps show qualitatively similar features on the ETO and STO surface states: a circular ring centered on Γ and two ellipsoids elongated in $k_x$ and $k_y$ directions (inset of Fig. 1d). Both structures are assigned to bands originating from $t_{2g}$ Ti-3d states, the ring-shaped one corresponding to light effective mass electrons having main $3d_{xy}$ orbital character, and the ellipsoidal ones related to heavy effective mass electrons with main $3d_{xz}$-$3d_{yz}$ characters. We observe notable differences between 2 uc and thicker ETO films, which are reflected also in the band dispersions (Fig. S3 in Supplementary Information), suggesting a contribution in 2uc ETO from the underlying STO. Thus in the following we focus our analysis on 15uc ETO, and in particular on 15uc ETO deposited on $SrRuO_3$ which has the best signal to noise ratio.

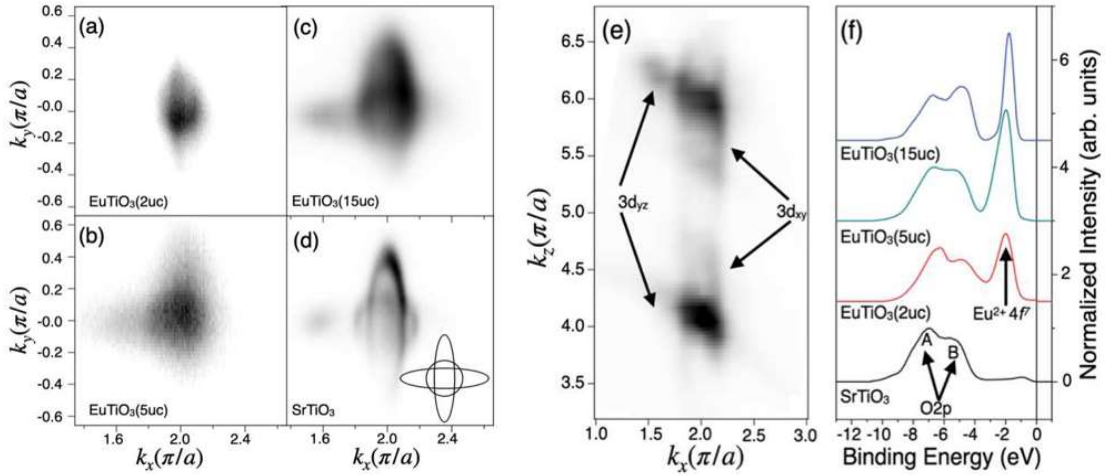

**Fig. 1**: (a)-(d) FS in the second Brillouin zone measured with an incoming photon energy of 85 eV and C+ polarization: (a) 2uc ETO, (b) 5uc ETO, (c) 15uc ETO and (d) STO samples. In the inset of (d) we show a sketch of the FS based on independent $t_{2g}$-states bands. e) $k_x$-$k_z$ cut measured on the 15uc ETO showing the two-dimensional dispersing $3d_{xy}$ light band and the more 3D-like $3d_{yz}$ heavy band. (f) Angle integrated VB spectra on STO and ETO based q2DEGs. Features A and B indicate the O-$2p$ band. A gradual inversion between the intensities of the A and B peaks going from STO to ETO films is related to the differences in the Sr-O-$2p$ and Eu-O-$2p$ hybridization in the two materials.

In Fig. 1e we show a $k_x$-$k_z$ plane cut at $k_y = 0$ measured on the 15uc ETO film by changing the incoming photon energy. The Fermi sheet corresponding to the light band has a nearly cylindrical shape with a $k_z$-independent 2D-character typical of a q2DEG, while the heavy bands show a $k_z$ dispersion related to a more 3D-like character. These findings, similar to those observed on the (001) STO surface state [15], demonstrate that ETO(001) hosts a q2DEG at its surface.

On the other hand, (001) ETO and STO surface states show important distinctive features. Fig. 1f shows the evolution of the angle-integrated valence band (VB) spectra from reference STO q2DEG to (001) ETO thin films of different thicknesses. The data show in ETO the emergence of a $Eu^{2+}$ narrow band, with a non-dispersing character (Fig. S4 in Supplementary Information), 1.95 eV below the Fermi level, related to localized $4f^7$ states.

To evaluate the main differences between the ETO and STO q2DEGs, in Fig. 2 we compare high statistics band dispersion, $k_x$-cuts through Γ point, acquired with different linear polarizations on 15uc (001) ETO film (Figs. 2a-b) and on two (001) STO single crystals (Figs. 2c-d STO-A, Figs. 2e-f STO-B) having 2D carrier densities $n_{2D}$ of $1.1 \pm 0.1 \times 10^{14}$ cm$^{-2}$ (STO-A) and $1.4 \pm 0.1 \times 10^{14}$ cm$^{-2}$ (STO-B), slightly below and above that of ETO ($n_{2D} = 1.3 \pm 0.1 \times 10^{14}$ cm$^{-2}$). The band dispersion profiles were determined by fitting the intensity maxima of the momentum dispersion curves (MDCs) at different binding energies (Fig. S5 in Supplementary Information); $n_{2D}$ were estimated from the Luttinger volume of each band.

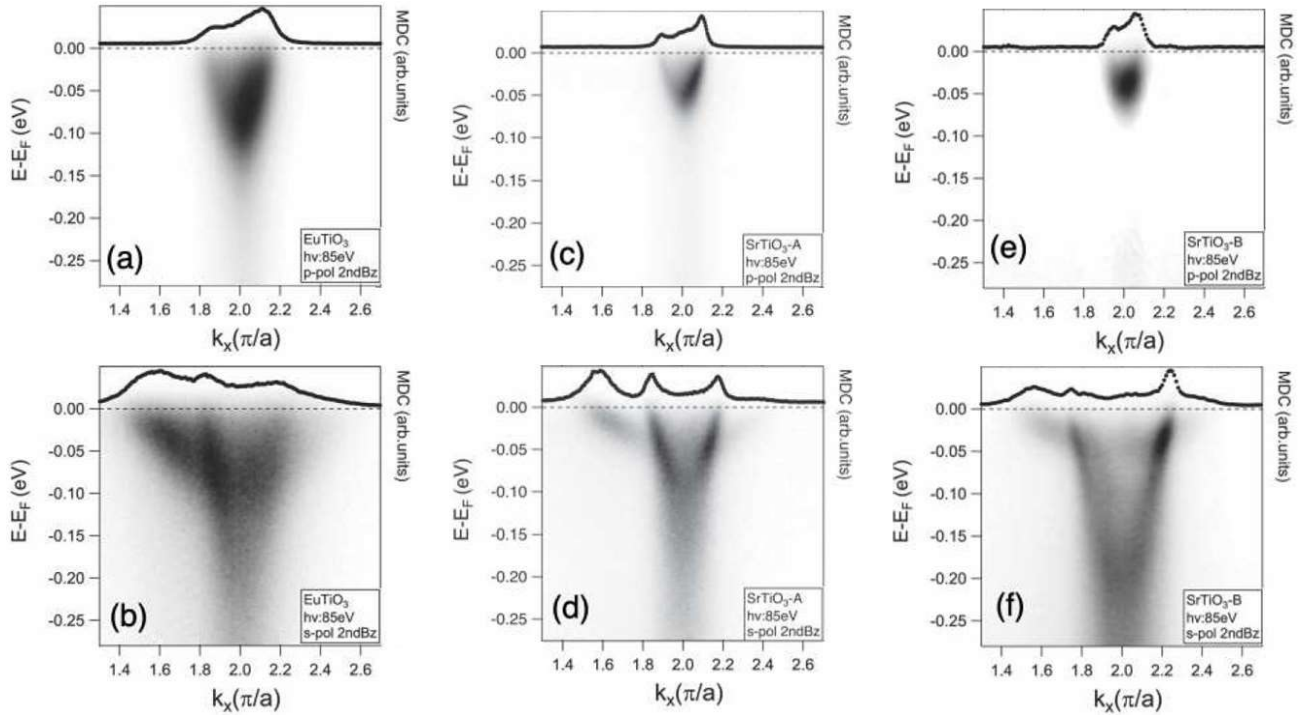

**Fig. 2**: High resolution dispersion maps on STO and ETO (001) surfaces: second Brillouin zone $k_x$ dispersion maps measured on 15uc ETO (a-b), STO-A (c-d) and STO-B q2DEGs (e-f), with incoming photon energy of 85 eV. The corresponding MDCs at the Fermi level, integrated over an energy window of 10 meV around the Fermi level, are shown as traces (circles) on the top of each panel.

Through dipole selection rules we identify the bands in the *p-pol* maps of Fig. 2 (left panels) as the $3d_{xz}$ band observed along the "narrow" k-diameter of the ellipse, while the bands observed in the *s-pol* maps (right panels) are light $3d_{xy}$ bands and heavy $3d_{yz}$ bands along the "wider" k-diameter of the ellipse. Additional features in MDCs at the Fermi level suggest the presence of at least one additional light band located above the lowest $3d_{xy}$ band in both ETO and STO-B samples. The data show that the band bottom of the $3d_{xz}$ and $3d_{yz}$ heavy bands in the ETO q2DEG is substantially lower than the one measured in the two STO q2DEGs, independent on $n_{2D}$. Instead, the $3d_{xy}$ band bottom in ETO is lower than the one of the STO-A and remarkably higher than in STO-B (Fig. 2d), where it reaches a value of -210 meV. Consequently, the splitting at the Γ-point between the low lying heavy and light bands in ETO, of the order of 60 meV, is smaller than the one typically measured in high carrier density STO q2DEGs and in other titanates like $CaTiO_3$ [18]. A simple tight binding interpolation of the band features, assuming independent $3d_{xy}$, $3d_{xz}$ and $3d_{yz}$ bands, furnishes the effective masses in the different samples (Fig. S6 in Supplementary Information). As shown in Table I, the ETO q2DEG is characterized by lower effective masses, compared to the STO surface states. Moreover, the band dispersions substantially deviate from the simple non-interacting tight-binding scenario for both STO and in particular ETO q2DEGs at the anti-crossing point and near the Fermi level (Fig. S7 in Supplementary Information). This implies a hybridization between $t_{2g}$ bands with different symmetries, induced by crystalline distortion from octahedral environment and by the SOC, which naturally mixes the bands: consequently, while retaining some of the original orbital character, ETO conduction band electrons are characterized by coupled spin and orbital degrees of freedom, which could induce an exotic spin-orbit texture, with both in-plane and out-of-plane components. This affects the general charge and spin magneto-transport properties.

Table I: Effective masses (in units of electron mass) of the ETO and STO q2DEGs obtained from tight-binding fit of the band dispersions and comparison with DFT+*U* calculations.

| | From tight binding fit | | | | DFT+*U* | |
|---|---|---|---|---|---|---|
| | $m_{xy}$ ($k_F$) [±0.05] | $m_{xy}$ (0) [±0.05] | $m_{xz}$ ($k_F$) [±0.05] | $m_{yz}$ ($k_F$) [±1] | $m_{xy}$ (0) | $m_{yz}$ ($k_F$) |
| ETO | 0.50 | 0.40 | 0.30 | 19 | 0.5 | 17.1 / 19.0 (AFM / FM) |
| STO-A | 0.60 | 0.50 | 0.50 | 23 | 0.6 | 22.2 |
| STO-B | 0.70 | 0.60 | 0.50 | 28 | | |

Since the conduction bands in ETO and STO are both formed by Ti-3*d* states, the question arises why the two q2DEGs show these differences and if the magnetic ordering of $Eu^{2+}$, demonstrated by magnetization data, could induce a spin-polarization in the q2DEG. For these reasons we performed DFT+*U* calculations of STO and ETO (001) surfaces using the Vienna ab-initio simulation package (VASP) [19, 20] with the projector augmented wave (PAW) basis [21, 22]. PBEsol exchange correlation functional [23] along with on-site Hubbard terms *U*=2eV for Ti-3*d* and *U*=7.5eV for Eu-4*f* states as implemented within the Dudarev's scheme [24] were used. The *U* value for Eu-4*f* is common for Eu-containing compounds and reproduces the position of the experimental $Eu^{2+}$ peak in VB, while for Ti-3*d* we used the typical value from literature [25, 26]. The main results of the calculations are not strongly affected by the choice of *U* for Ti-3*d* states, although some details concerning the band dispersion and the position of the O-2*p* bands depend on this choice. The ionic positions were fully relaxed until the forces were less than 0.001 eV/Å.

Core-level spectroscopy suggests that ETO films are TiO$_2$ terminated (Supplementary Information). Thus, we show here DFT+$U$ results for TiO$_2$ terminations, while in Supplementary Information we report calculations for EuO termination.

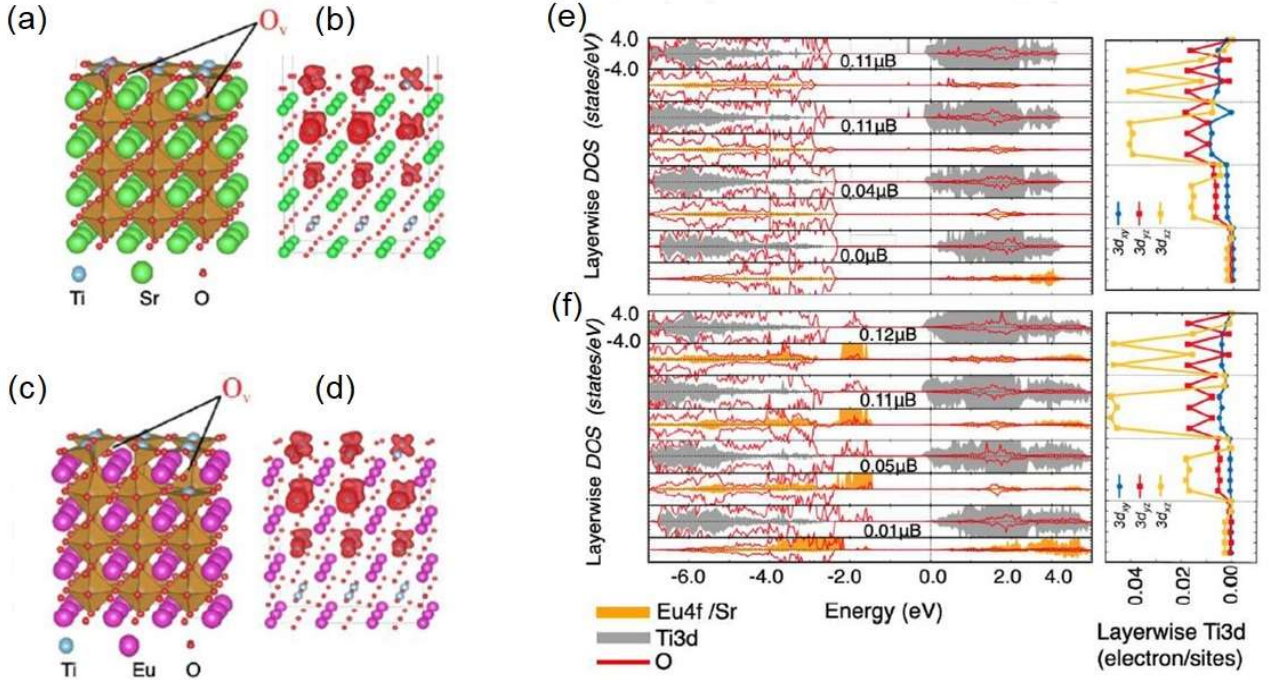

**Fig. 3**: Layer-, element- and spin resolved density of states of TiO$_2$-terminated ETO(001) and STO(001) surfaces from the DFT+$U$ calculations: side views of the relaxed (a) STO(001) and (c) ETO(001) surfaces with an oxygen divacancy in the topmost TiO$_2$ layer and a second one in the subsurface SrO and EuO layers, respectively. Panels (b) and (d) show the spin density (integrated between -0.4 eV to Fermi level) with iso-values of 0.0005 e/Å$^3$. The color scale (blue to red) indicates predominant majority (red) spin contribution. (e) STO and (f) ETO layer-, spin- and element- resolved density of states together with orbital-resolved Ti-3$d$ occupation (electrons per site) showing the spatial distribution, the orbital polarization and Ti- spin-magnetic moment of the q2DEGs.

Since ideal surfaces are insulating, we considered models containing an oxygen divacancy in various planar and vertical configurations, inducing the formation of a q2DEG [27, 28]. In agreement with previous studies on STO (001) [27], we find that the most energetically favored configuration for both surfaces is a divacancy in a 2$a$ × 3$a$ × 4$a$ supercell ($a$ =3.905°A), with one missing oxygen in the topmost TiO$_2$ layer and a second in the subsurface Sr(Eu)O layer at a distance of 6.2 Å (Figs. 3a and 3c). In the following, we show mostly results of the calculations performed using this configuration. Calculations of different magnetic configurations indicate that the FM solution is the most stable in both (001) STO and ETO, giving a localized Ti- magnetic moment at the surface of about 0.12 $\mu_B$/Ti. Figs. 3b,e (STO) and 3d,f (ETO) show the spatial distribution of the q2DEG and of the Ti-spin-moment, together with the layer-, spin- and element-resolved density of states (DOS) of the two systems. The O-2$p$ VB extends from ~ -7 to -2.5 eV in both systems, while the spin polarized Eu-4$f^{\,7}$ orbitals appear in ETO (001) as a narrow (0.7 eV wide) band around -2 eV, in excellent agreement with experimental results. Above -1.5 eV the DOS contains contribution with predominant Ti-3$d$ character and comprises the coexistence of localized states at -0.8 eV, in STO(001) surface only, as well as delocalized states in both ETO and STO surfaces contributing to the q2DEG.

To highlight the differences between the electronic band structure of (001) STO and ETO surface states, we show in Fig. 4a-c and 4d-e the spin-resolved theoretical band dispersions, calculated within the planar and vertical oxygen divacancy configurations, respectively, without including SOC. Further calculations including SOC are shown in the Supplementary Information. DFT+$U$ show that several bands cross $E_F$, the lowest lying one having a dispersive $d_{xy}$ character switching along Γ−X direction into a much flatter band indicating an avoided crossing with a $d_{xz}$, $d_{yz}$ bands. The higher lying bands around Γ are a combination of $d_{xy}$ and $d_{xz}$, $d_{yz}$ bands, indicating replicas from different layers due to the q2DEG confinement. In both planar and vertical oxygen divacancy configurations, overall, the bands in ETO (001) reach deeper below $E_F$ compared to STO (001). The strong change in dispersion for the lowest lying band reflects in the effective masses around Γ (in both FM and G-AFM solutions, see Table I). In particular, within the FM planar oxygen divacancy configuration we find m*= 0.6 $m_e$ and 0.5 $m_e$ for STO (001) and ETO (001), respectively, vs. 22.2 $m_e$ and 17.1 $m_e$ for the flat part along Γ−X. These values are in reasonable agreement with those extracted from ARPES data, confirming the experimental evidence of lower effective masses for the q2DEG at the (001) ETO surface. Moreover, DFT+$U$ calculations correctly reproduce the overall downward shift of the heavy bands in the ETO (001) q2DEG. This downward shift is partly due to larger band-bending at the (001) ETO, due to the different lattice screening compared to (001) STO, and to a possible hybridization between $Eu^{2+}$ and the conduction band. Indeed, the $Eu^{2+}$ feature has a long tail merging with the quasiparticle peak, as shown in the energy distribution data (Fig. S4c in Supplementary Information).

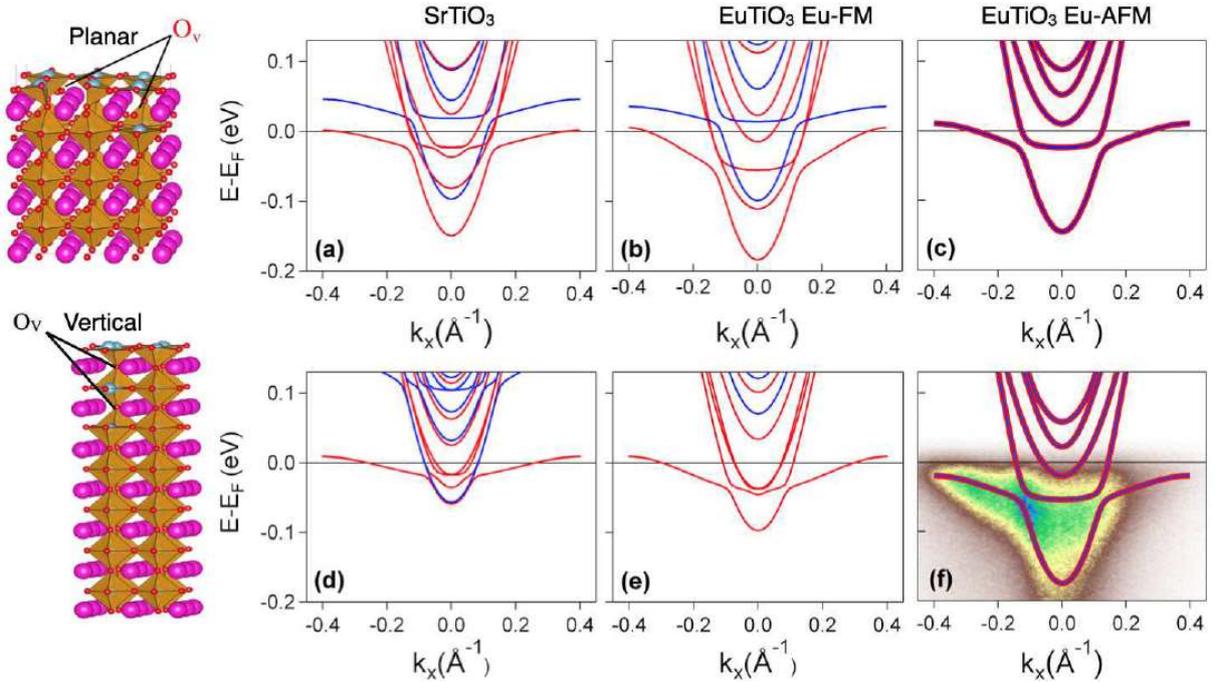

**Fig. 4**: Spin resolved band dispersions along Γ-X from the DFT+$U$ calculations in the planar (panels (a)-(c)) and vertical (panels (d)-(e)) oxygen divacancy configurations without spin-orbit coupling (SOC) in STO (001) (panels (a) and (d)) and ETO (001) (panels (b) and (c) and (e)). On the left we show also the supercells used in these calculations, with the location of the oxygen vacancies ($O^{2+}V$). For ETO (001), panels (b) and (c) show results with Eu sub-lattice in FM and AFM order, respectively, highlighting how the FM order is crucial for the spin splitting. (f) A comparison between DFT+$U$ in the AFM state of ETO (after a rigid band shift of -30 meV) and ETO band dispersion measured at 15K with s-pol polarization. In each panel red and blue lines are spin-up (majority) and spin-down (minority) bands.

In both systems a spin splitting between majority (red) and minority (blue) bands emerges, however it is larger in ETO: ~0.1 eV and ~0.2 eV, for planar and vertical divacancy configurations, respectively. In STO (001) the largest spin-splitting is ~0.05 eV for the planar divacancy configuration, while almost spin-degenerate bands are found in the vertical oxygen divacancy configuration. We remark that the oxygen vacancies in the supercell calculations are long-range ordered, thus a prediction of a FM ground state in some configurations also in STO (001) is not surprising. However, a 1 × 1 lattice periodicity indicates that the oxygen vacancies are not ordered. This suggests that a FM q2DEG at the STO (001) surface can be obtained only by properly engineering the oxygen vacancies, maybe explaining some controversies among recent spin-resolved ARPES studies [29, 30]. However, since ordered oxygen vacancies alone can lead to a spin-polarized q2DEG, one may wonder if the $Eu^{2+}$ ferromagnetism is really necessary. In order to understand the role of Eu magnetism in the spin-polarization of the q2DEG, in Fig. 4c we show that, within the G-AFM solution of the model, the Ti 3$d$ bands are not spin-polarized. The AFM solution can be regarded as a first approximation of the paramagnetic state and, after a small rigid downward shift of 30 meV, it is in good agreement with the ARPES dispersion measured at 15 K, thus above the FM $T_c$. A direct experimental confirmation of a spin- polarized q2DEG at the ETO (001) surface would require a spin-resolved ARPES experiment below the FM $T_c$ on a single domain of the FM-ETO surface state, which is unfeasible with the available setups. On the other hand, DFT+$U$ predictions, and all the reported data suggest that the (001) surface of ferromagnetic ETO hosts a spin-polarized oxide q2DEG, confirming recent experimental and theoretical studies on ETO [13, 14, 31] and (001) LaAlO$_3$/ETO/STO heterostructures [32]. In particular, DFT+$U$ calculations show that the exchange interaction between Ti-3$d$ states and ordered $Eu^{2+}$ magnetic moments in the lattice is essential to create a spin-polarized q2DEG. We anticipate that other systems where a similar spin-polarized 2DEGs could be realized are GdTiO$_3$ [33] and NdTiO$_3$ [34], which are the magnetic counterpart of the 3$d^1$ LaTiO$_3$ Mott insulator.

The results presented in this work establishes EuTiO$_3$ as a new material platform for oxide q2DEGs characterized by broken inversion and time reversal symmetries, of interest for the study of novel quantum phenomena in low dimensional systems.

This project received funding from the ERA-NET QUANTERA European Union's Horizon H2020 project QUANTOX under Grant Agreement No. 731473, from the Ministero dell'Istruzione, dell'Università e della Ricerca for the PRIN project TOP-SPIN (Grant No. PRIN 20177SL7HC). R.P. and M.V. acknowledge funding by the German Research Foundation (DFG) within CRC/TRR80 (project number 107745057, subproject G3) and CRC1242 (project number 278162697, subproject C02) and computational time at the Leibniz Rechenzentrum Garching, project pr87ro and supercomputer magnitUDE (DFG grants INST 20876/209-1 FUGG, INST 20876/243-1 FUGG).

# Supplementary Information

## of

## A two-dimensional electron gas at the (001) surface of ferromagnetic EuTiO$_3$

R. Di Capua, M. Verma, M. Radović, N. C. Plumb, J. H. Dil, Z. Ristić, E. B. Guedes, G. M. De Luca,
D. Preziosi, Z. Wang, A. P. Weber, R. Pentcheva, and M. Salluzzo.

**S1 - Deposition and structural characterization of EuTiO$_3$ films**

EuTiO$_3$ films were *in-situ* grown by Pulsed Laser Deposition (PLD) in 10$^{-6}$ mbar molecular oxygen partial pressure (base pressure 1x10$^{-9}$ mbar) at 720°C on 5x10x0.5 mm$^3$ TiO$_2$-terminated (001) SrTiO$_3$ (STO) single crystals using the facility available at the Surface/Interface Spectroscopy (SIS, X09LA) beamline of the Swiss Light Source (SLS) - Paul Scherrer Institute. The number of unit cells (uc) (from 2 to 15 uc) and the structural quality of the samples was monitored during the growth by *in-situ* reflection high-energy electron diffraction (RHEED) and after deposition using LEED (Low Energy Electron Diffraction). We have also deposited a 15uc EuTiO$_3$ sample on a SrRuO$_3$ buffer layer to improve contact to the ground. In Fig. S1a we report the RHEED intensity vs. time of the specular and first order diffraction spots during the growth of a nominal 15uc EuTiO$_3$ (ETO) film, and in Fig. S1b the final RHEED pattern. Regular oscillations are visible during the growth, while the final RHEED pattern show the high structural quality of the surface.

As grown samples are insulating, and are charging under the beam irradiation, thus they cannot be measured by ARPES unless after an unusual long exposition to the intense synchrotron radiation beam (up to 24 h). In order to get a conducting surface, each sample was annealed in Ultra High Vacuum (UHV) at 600 °C for one hour. This results in a metallic surface state, which was studied by ARPES. The annealing process does not affect the 1x1 surface structure and the in-plane lattice parameters, as demonstrated by the combination of RHEED and LEED measurement shown in Fig. S1b and Fig. S1c respectively. From RHEED (Fig. S1b) and LEED (Fig.S1c) data, we can accurately measure the in-plane lattice parameters as function of the ETO thickness, which are found to be identical to that one of the STO substrate, thus perfectly matched with the STO in-plane lattice.

In order to further characterize the samples structural properties, we performed ex-situ x-ray diffraction measurements using a lab-based diffractometer at the University of Strasbourg (Rigaku Smartlab (9 kW) diffractometer in parallel beam geometry equipped with a Cu kα1 source (kα1 = 1.5406 A°) and a Ge(220) monochromator). The perfect in-plane matching between the ETO overlayer and the STO substrate is confirmed by reciprocal space mapping x-ray diffraction data around the (103) diffraction peak, as shown in Fig.S1d for the ARPES measured 15 uc ETO sample. From these data we can also see that the (103) diffraction peak of STO and ETO cannot be separated, suggesting a similar c-axis.

In Fig. S1e, we show θ−2θ x-ray diffraction on the same sample. In Fig. S1f we also show data for thinner ETO films, i.e. an 8uc ETO protected by a LAO epitaxial overlayer, grown in the same deposition conditions. The LAO capping preserves the valence of Eu, which on the other hand changes at the surface from Eu$^{2+}$ to Eu$^{3+}$ if exposed to moderate oxygen pressure or to ambient atmosphere.

Both samples show Pendellösung size effect fringes at the left and at the right of the STO (00L) peaks, which are related to the finite thickness of overlayer. The main ETO (00L) peaks are not separated at these low thicknesses from the STO Bragg peaks, however we can get quite precise information about the ETO c-axis from a fitting of the (001) data using dynamical x-ray diffraction code as implemented in the software Gen-X, widely used to analyze data for thin epitaxial films and multilayers [S1]. We modeled the systems as the combination of the STO bulk (TiO$_2$ terminated), and nominal ETO (and LAO for the heterostructure) layer thicknesses, as inferred from the RHEED oscillations during the growth. The obtained fits are very accurate (R-factor<=.2) and are very sensitive to the c-axis of the ETO (and LAO) layer, as the periodicity of the Pendellösung fringes depends on them). For the non-capped ETO film we find a c-axis of 3.92± 0.01 Å, and for the capped film a c-axis of 3.91± 0.02 Å for ETO and of 3.790± 0.005 Å for LAO (the error in the estimation of the c-axis for the heterostructure is larger and is determined from the sensitivity of the fit to this parameter). A slight elongated c-axis in the non-protected ETO film suggests a higher oxygen vacancy concentration, due to the change in the oxidation of Eu under air exposition.

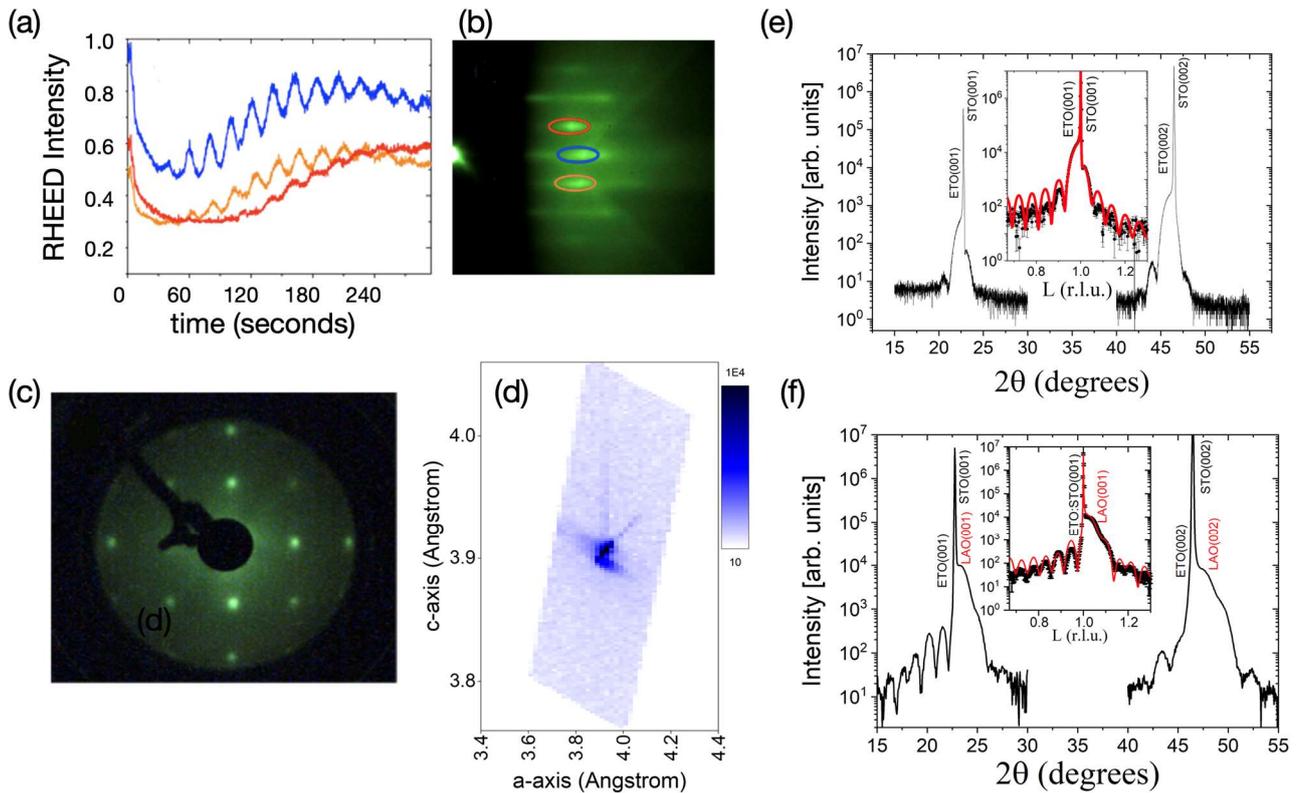

**Fig. S1**: a) Real-time intensity oscillation of the specular (blue) and first order (1 0) and (-1 0) (orange and red curves) diffraction spots in the RHEED pattern during the growth of a 15uc ETO films. b) Corresponding RHEED pattern at the end of the deposition. The ellipses represent the regions used to integrate the intensity shown in a) with their corresponding color code. c) LEED pattern, on the same film, after the post-annealing process. (d) ex-situ x-ray diffraction reciprocal space x-ray diffraction map around the (103) reflection. (e) and (f) θ-2θ pattern on the 15 uc ETO/STO film measured by ARPES and of a LAO(11uc)/ETO(8uc)/STO heterostructure grown in the same conditions, respectively. Both data show Pendellösung fringes around the main diffraction peak. In the insets we show the (001) diffraction peaks and their fit (red lines).

## S2 – Ferromagnetic characterization of EuTiO$_3$ (ETO) films

We performed Eu $M_{4,5}$ and Ti $L_{2,3}$ edges x-ray magnetic circular dichroism (XMCD) measurements in total electron yield (TEY) mode on representative samples protected (capped) or not-protected (uncapped) by a LaAlO$_3$ overlayer, namely LaAlO$_3$/EuTiO$_3$/STO heterostructures and EuTiO$_3$ thin films. In particular, in Fig. S2, we show XMCD data at 6.5 and 0.1 Tesla on 8uc (3nm) EuTiO$_3$ films (Fig.S2a-c) and on a similar sample covered by a 4 nm LAO epitaxial film (Fig. S2d-f), grown in the same conditions of the samples measured by ARPES.
The comparison between capped and uncapped ETO shows a sizeable XMCD-signal at the Eu $M_{4,5}$ and Ti-$L_{2,3}$ edges in both cases. However, only the capped sample is characterized by XAS and XMCD spectra typical of Eu$^{2+}$ oxidation state, with only a minor percentage of Eu$^{3+}$, and an Eu spin moment close to the value expected for Eu in Eu$^{2+}$ valence state. It is also clear from the data that in the large part of the uncapped (and air exposed) ETO, Eu-ions change valence from Eu$^{2+}$ to Eu$^{3+}$ (Fig. S2a). As the probing depth in TEY at 1100 eV is of the order of 8-10 nm (and 2 nm at the 460 eV), the valence change happens at the surface and in a substantial fraction of the investigated sample (3 nm thick). These data show that in uncapped samples after exposure to air, Eu$^{2+}$ valence changes to Eu$^{3+}$ in a substantial fraction, which explains the reduced XMCD signal. Yet, the data show a clear Ti-$L_{2,3}$ edge XMCD even in uncapped ETO at 0.1 Tesla, although it was air-exposed, showing that ETO has a finite splitting of Ti-3d states.
We also performed SQUID magnetometry on the capped LAO/ETO sample and on the unprotected ETO/STO 15uc film measured by ARPES. As shown in suppl. Fig. S2g), a clear hysteresis in the magnetization at low temperature (3 and 4 K), disappearing at 10 K is observed. In protected ETO, as expected, a total saturated magnetic moment of about 6.8 m$_B$/Eu, close to the one expected from Eu$^2$, is found, while the unprotected ETO show a reduced saturation magnetization of about 4.0 m$_B$/Eu due to the presence of non-magnetic Eu$^{3+}$. These results demonstrate unambiguously that EuTiO$_3$ films are ferromagnetic, with a Tc below 10 K.

*Supplementary Information*

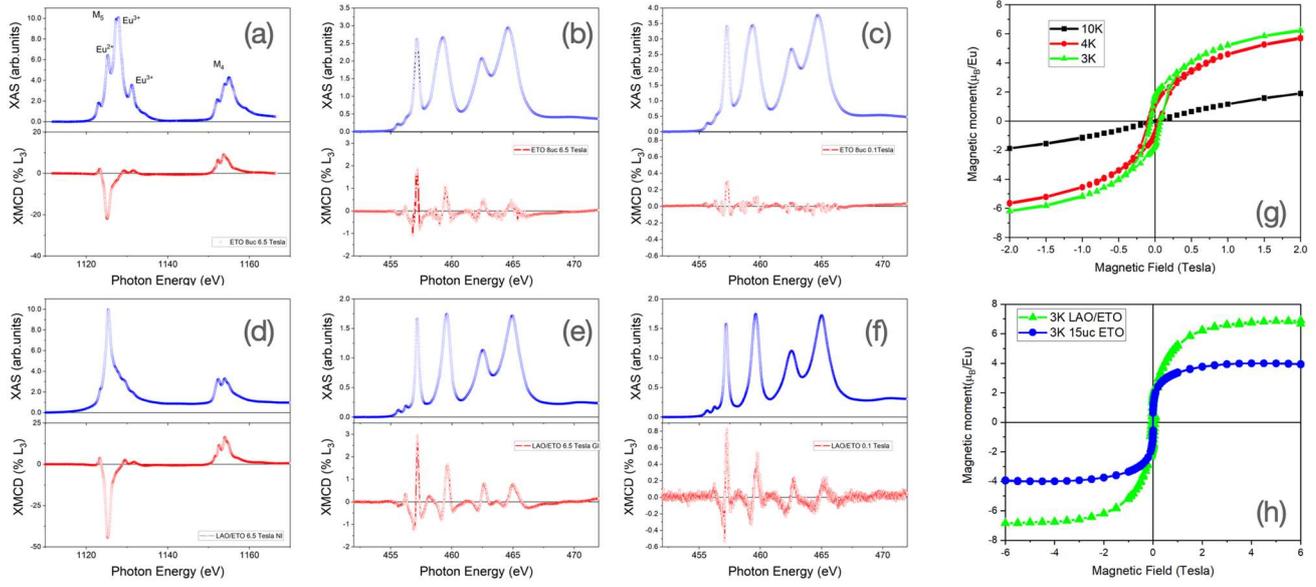

**Fig. S2:** Eu-$M_{4,5}$ edge and Ti-$L_{2,3}$ edge XAS and XMCD data on uncapped (a)-(c) and LAO-capped (d)-(f) ETO films. (a)-(b) and (d)-(e) spectra are measured at 6.5 Tesla, while (c) and (f) Ti-$L_{2,3}$ edge XAS and XMCD data are measured at 0.1 Tesla. The data are acquired in grazing incidence (70 degrees from the surface normal) in total electron yield mode at 3K.(g) SQUID magnetometry on LAO capped 8uc ETO/STO sample at three temperatures, 3K, 4K, and 10K, and (h) comparison on large field range with SQUID magnetometry data on 15uc ETO. The well-known diamagnetic contribution of the STO substrates has been subtracted from the data.

**S3 Supplementary dispersion maps as function of the ETO thickness**

We show in Fig. S3 the evolution of the band dispersions of ETO thin films as function of the thickness i.e., 2, 5 and 15 uc ETO deposited on insulating, $TiO_2$ terminated, STO single crystals and of 15uc ETO deposited on a buffer $SrRuO_3$ metallic layer.
From the thickness evolution, we notice that 2uc ETO shows a band dispersion quite different from thicker ETO films. In particular the band bottom is around -60 meV much higher than that one of thicker ETO, (-160 meV), and the splitting between heavy and light bands is very small (if any).
This result suggests that the in 2uc ETO the surface state is not fully developed and that we do see contributions from the STO bulk. At 15 uc, ARPES data of the sample deposited on bare insulating STO and on SRO buffer are quite similar, with similar Fermi momentum and band bottom. As expected, the signal to noise ratio is much higher in the sample deposited on conducting SRO.

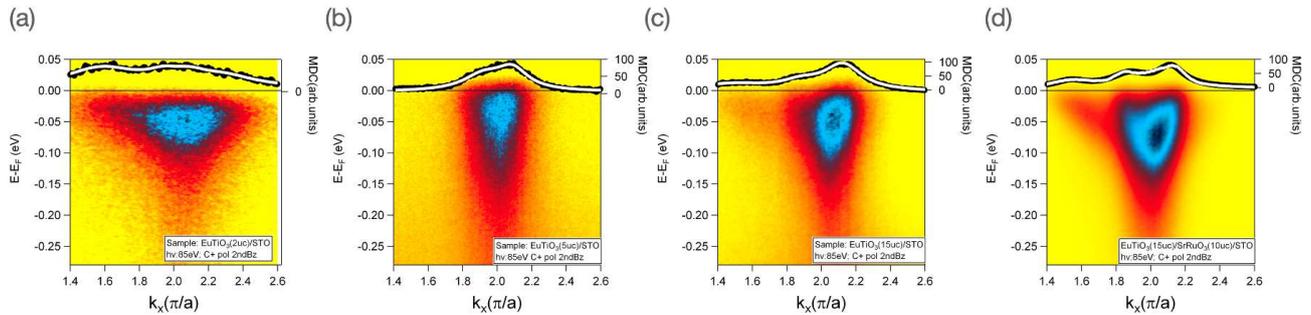

**Fig. S3**: Evolution of the band dispersion measured with hν = 85 eV and C+ polarization: (a) 2uc ETO on STO; (b) 5uc ETO on STO; (c) 15uc on STO; (d) 15 uc ETO on SRO buffer layer deposited on STO.

*Supplementary Information*

In Table S1 we compare the values of the Fermi momentum, the band bottom for heavy and light bands estimated from the analysis of MDC-curves at the Fermi level and at several energies below $E_F$, combined to analysis of the 2D-curvature maps of the data ([S2]).

**Table S1**: values of the Fermi momenta of the different bands identified in the band dispersion cuts. For the 5uc sample the value of $k_F$ for heavy bands cannot be determined from the dispersion map, being too week. On the other hand, signatures of heavy bands are seen in the Fermi surface data shown in Fig.1 of the main text. E(0) is the band bottom.

| Sample | $k_F$ (Å$^{-1}$) $3d_{xy}$ | $k_F$ (Å$^{-1}$) $3d_{xz}$ | $k_F$ (Å$^{-1}$) $3d_{yz}$ | $-E_0$ (meV) Light band | $-E_0$ (meV) Heavy bands |
|---|---|---|---|---|---|
| 2uc ETO/STO | 0.08±0.01 | 0.08±0.01 | 0.40±0.01 | 60±5 | 60±5 |
| 5uc ETO/STO | 0.07±0.01 | - | - | 80±5 | - |
| 15uc ETO/STO | 0.13±0.01 | 0.08±0.01 | 0.37±0.01 | 150±5 | 90±5 |
| 15ucETO/SRO/STO | 0.135±0.005 | 0.090±0.005 | 0.35±0.01 | 160±5 | 100±5 |

**S4-Overview of the valence band dispersion of ETO q2DEG**

In Fig. S4, we show the valence band $k_x$- dispersion map (a), and its corresponding curvature map (b) (ref. [S2]) on a large binding energy range acquired with 85 eV incoming photons and C+ polarization around the Γ-point. One can see the almost non dispersing character of the Eu-4f band, in contrast with the more dispersing character of O-2p bands.
In Fig. S4c we show an energy dispersive curve average over -0.5 to 0.5 Å$^{-1}$ momentum around the gamma point showing the quasiparticle peak and the long tail of the Eu$^{2+}$ 4f$^7$ peak at -2 eV.

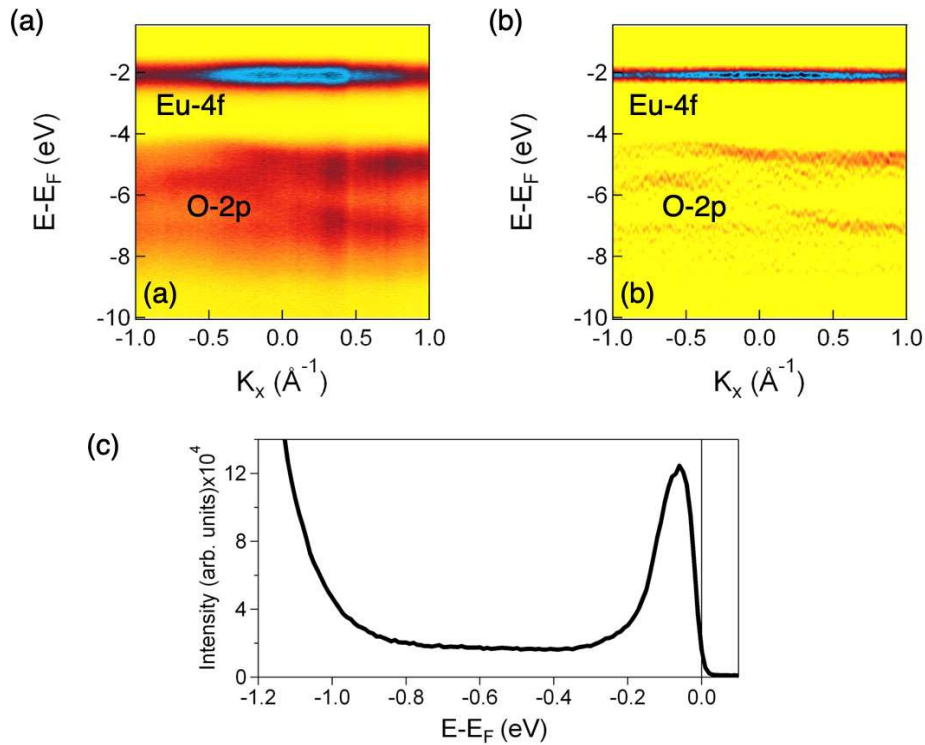

**Fig. S4** a) Valence band dispersion of ETO q2DEG showing the Eu-4f and the O-2p bands and (b) corresponding curvature map. (c) Energy dispersion curve average over -0.5 to 0.5 Å$^{-1}$ momentum around the gamma point, which show the quasiparticle peak merging with the long tail of the -2eV Eu4f band.

*Supplementary Information*

**S5 - MDC analysis.**

Momentum Dispersion Curves (MDCs) allow to localize maxima of ARPES signal in the E-$k_x$ dispersion maps, for the identification and the attribution of the recorded band features. In Fig. S5 we report examples of MDC profiles analysis on a 15uc ETO/SRO/STO sample, at different binding energy values $E$ (referred to the Fermi level $E_F$ as $E - E_F$). The MDC profile corresponding to a given $E$ value is obtained by averaging ARPES intensity values at each $k_x$ over an energy window of 10 meV (centered on the chosen $E$). The analysis of each profile was performed through a deconvolution in terms of overlapping Lorentzian or Gaussian peaks (both of them providing similar fittings). In the reported maps, the light $d_{xy}$ and the heavy one along the largest $k_F$ value, $d_{yz}$ emerge from the map recorded in s-polarization of incoming photons, while the $d_{xz}$ band (heavy band along the direction of minimum $k_F$) can be observed in the p-polarization. At each given $E$ value, the central $k_x$ of each Lorentzian/Gaussian contribution defines the ($k_x$, $E$) point to be employed in the tight binding fitting procedure described in the following.

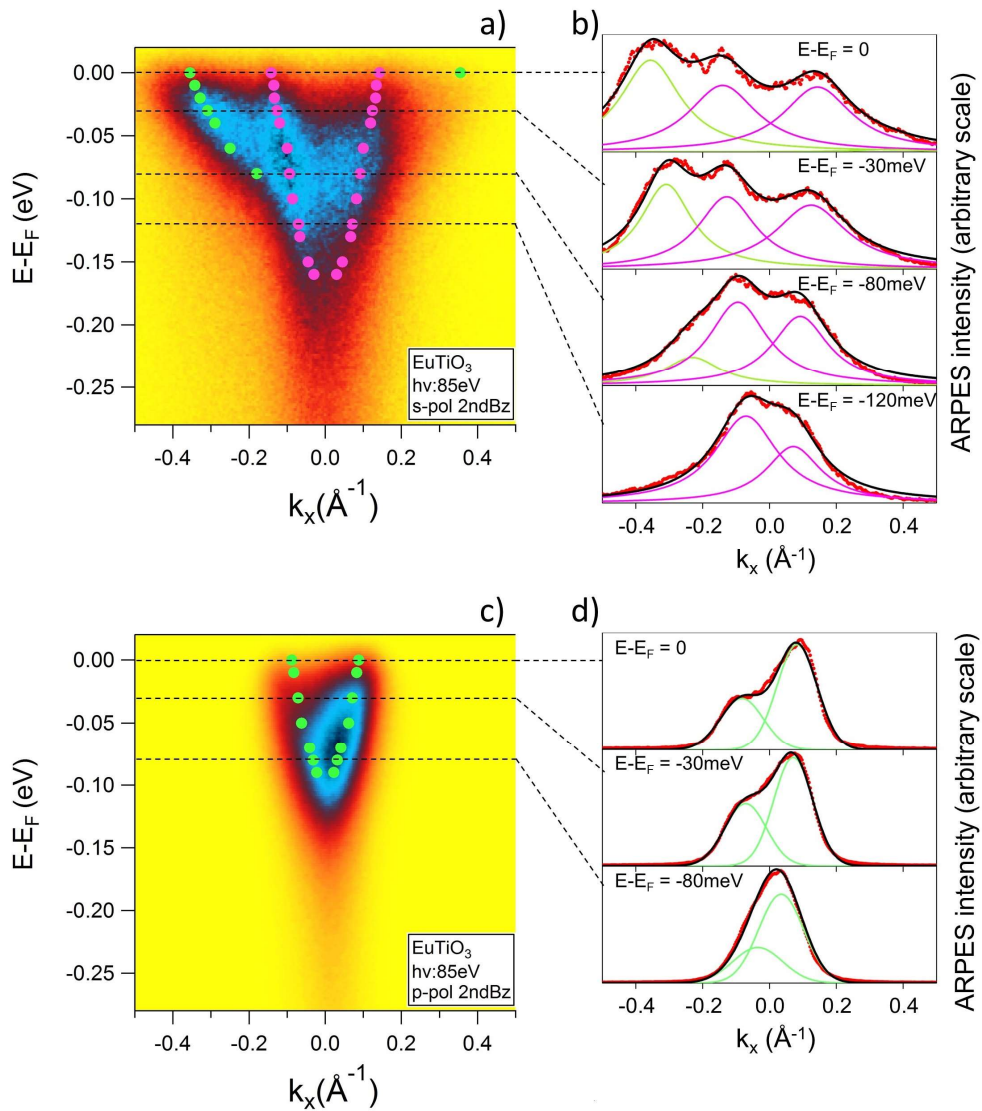

**Fig. S5**: High statistics band dispersion $k_x$-cuts (acquired around the Γ point in the second Brillouin zone, s- and p-polarization, cfr. main text) on the surfaces of the 15 uc ETO/SRO/STO film: example of MDC analysis for the extraction of ($k_x$, $E$) point to be employed in the tight binding fitting procedure described in the following. a) s-pol map, on which the points for the fitting are highlighted ($3d_{yz}$ as green circles, $3d_{xy}$ as magenta circles). b) MDC profiles for some of the binding energy values, with the peak deconvolution which leads to the identification of the points highlighted in panel a (contributions from the different bands are in the same color code as in panel a; the global fitting curve is reported as black line, the experimental data as red circles). c) p-pol map, on which the points for the fitting are highlighted ($3d_{xz}$ as green circles). d) MDC profiles for some of the binding energy values, with the peak deconvolution which leads to the identification of the points highlighted in panel c (the global fitting curve is reported as black line, the experimental data as red circles). For all the polarizations on all the investigated sample, we applied this procedure to infer the experimental points for the tight binding band fitting.

*Supplementary Information*

**S6- Tight binding band fitting**

In Fig. S6, we show the high statistics band dispersion $k_x$-cuts (around the Γ point in the Second Brillouin zone, cfr Fig. 2 main text) on 15uc ETO film, STO-A and STO-B samples, and the corresponding Fermi surfaces ($k_x$-$k_y$ cuts).

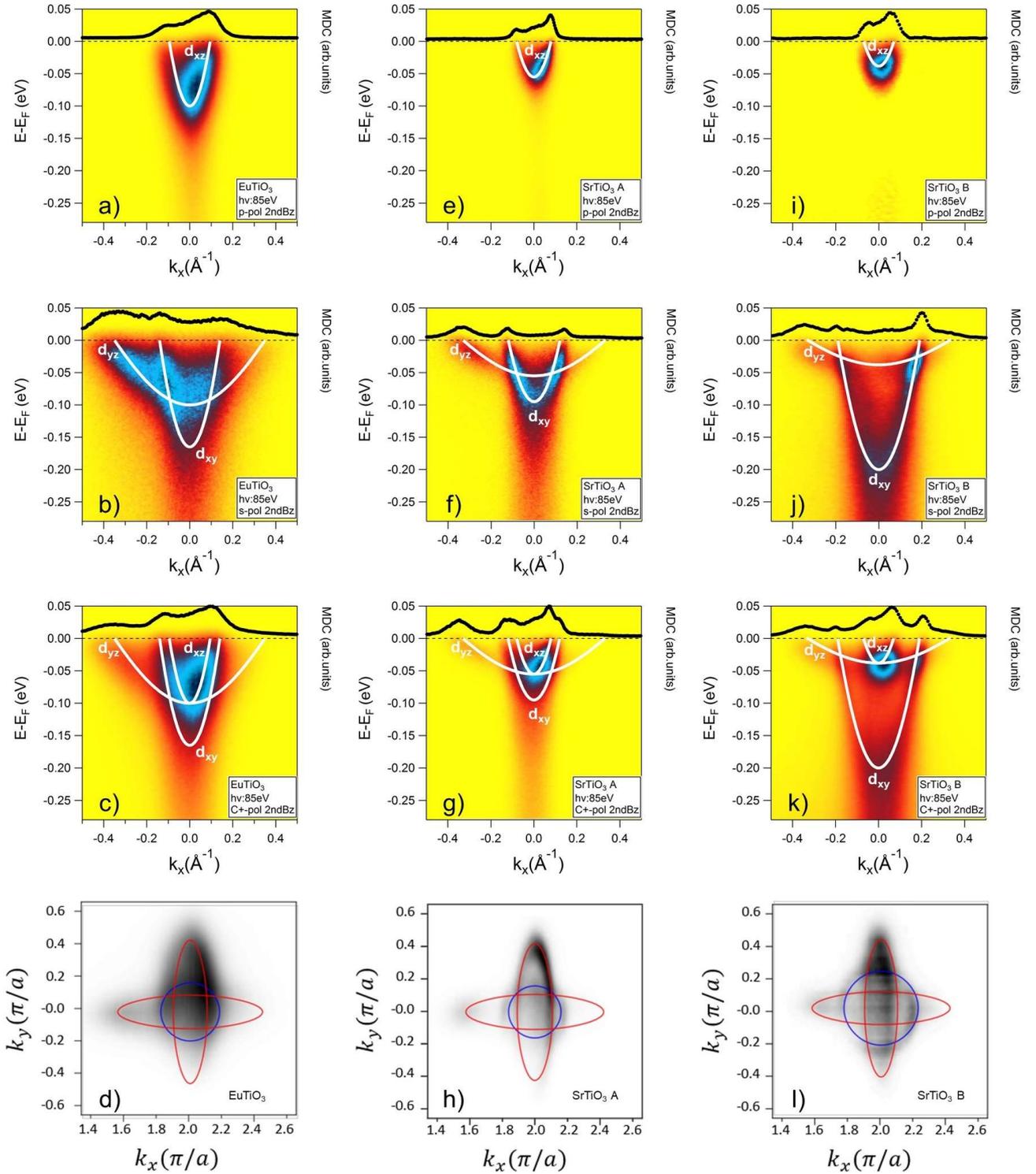

**Fig. S6**: High statistics band dispersion $k_x$-cuts and the relative $k_x$-$k_y$ cuts of the Fermi surface (data acquired around the Γ point in the second Brillouin zone), together with the corresponding tight binding fits. The panels report: band dispersions for the 15 uc ETO film in p-polarization (a), s-polarization (b), C+-polarization (c), and the Fermi surface of the 15 uc ETO film (d); band dispersions for the STO-A sample in p-polarization (e), s-polarization (f), C+-polarization (g), and the Fermi surface of the STO-A sample (h); band dispersions for the STO-B sample in p-polarization (i), s-polarization (j), C+-polarization (k), and the Fermi surface of the STO-B sample (l). (in the Fermi surface plots, the blue circumference is the cut of the $d_{xy}$ sheet, the red ellipses are the cuts of the $d_{xz}$-$d_{yz}$ sheets)

*Supplementary Information*

Here we show p-, s-, and C+-polarization maps, in false color scale, and the tight binding fit for the band identification. The performed tight binding fits were then used to estimate the Fermi momentum and the effective masses of electrons in each band of the three samples. (see Table 1 in the main text). Circular polarization (C+) $E$ vs. $k_x$ band dispersions include features associated to all the bands in the system, namely bands with $3d_{xz}$, $3d_{yz}$ and $3d_{xy}$ characters; dataset recorded in *p-pol* configuration are mainly affected by the $3d_{xz}$ band, while *s-pol* configurations provide information about the $3d_{xy}$ and the $3d_{yz}$ bands. ($d_{xy}$ indicates the light band, while $d_{xz}$ and $d_{yz}$ refer to the heavy band along the short and the long diameters of the ellipse, respectively). The band-dispersion profiles at the different polarizations were analyzed considering simple 2D tight binding dispersion relations [S3] taken at $k_y = 0$ as a function on only $k_x$. Assuming three independent bands, the fitting equations can be written in the following form:

$$E_{xy}(k_x) - E_F = V_{xy}(1 - cos k_x a_0) + E_{0xy}$$

$$E_{xz}(k_x) - E_F = V_{xz}(1 - cos k_x a_0) + E_{0xz}$$

$$E_{xy}(k_x) - E_F = V_{yz}(1 - cos k_x a_0) + E_{0yz}$$

Where $E_F$ is the Fermi energy, $a_0$ is the lattice constant of the cubic cell, $V_{xy}$, $V_{xz}$, $V_{yz}$ are related to the so-called inner potentials, and $E_{0xy}$, $E_{0xz}$, $E_{0yz}$, being the $E - E_F$ value at $k_x = 0$, represent the band bottom ($E_{0xz}$ and $E_{0yz}$ are expected to have the same value inside the experimental error).

## S7- Curvature maps
.

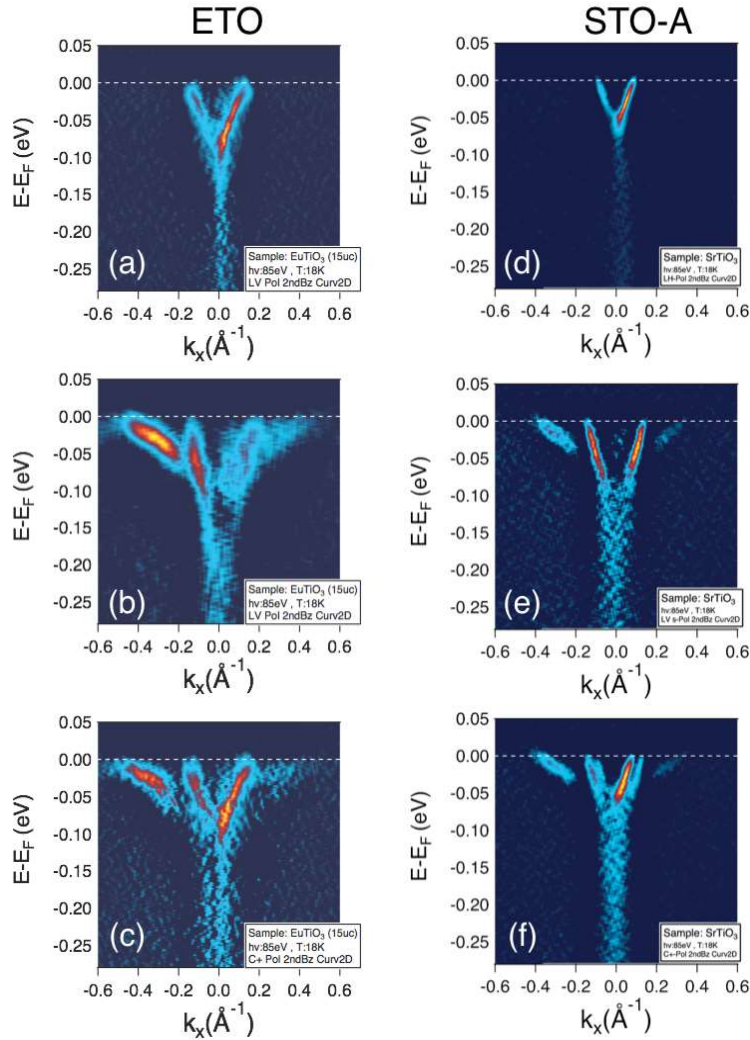

**Fig. S7**: High statistics 2D-curvature band dispersion maps for: a)-c) a 15 uc ETO film and d)-f) STO-A q2DEG. Panels a) and d) are for LH (p-pol) polarization, b)-e) for LV (s-pol polarization) and c) and f) for C+ polarization.

*Supplementary Information*

We applied the method described in ref. [S2] to extract 2D-curvature maps of the ARPES dispersions shown in Fig. 2 of the main text a to highlight the differences between ETO and STO electronic structure.
The curvature maps in Fig. S7 reveal that for both STO and ETO surface states the dispersion maps substantially deviate from independent bands modeled within a tight-binding scenario near the anti-crossing point and near the Fermi level, as confirmed by DFT+U calculations. In the ETO case, the deviation is more pronounced compared to the STO cases.

**S8- Surface termination of EuTiO$_3$ films: XPS core level spectroscopy of Eu and Ti states, and DFT+U calculations on EuO terminated surface.**

The main DFT+U calculations presented throughout the manuscript consider EuTiO$_3$ surfaces terminated with TiO$_2$. Here we provide experimental evidence and further DFT+U calculations, which suggests that the assumption is correct, i.e. our EuTiO$_3$ are TiO$_2$ terminated. First, we show in Fig. S8a,b core level spectroscopy data of the 15 uc ETO/SRO sample on which most of the analysis was focused on. In order to have info about the surface terminations, we used 600 eV energy photons to measure at the same time Ti-*2p* and Eu-*4d* core level spectra.
Eu-*4d* core level spectral features shown in Fig.S8a are typical of Eu in Eu$^{2+}$ oxidation state. On the other hand, Ti-*2p* core level spectra are characterized by main Ti-*2p$_{1/2}$* and *2p$_{3/2}$* (Ti$^{4+}$) peaks, and shoulders at lower binding energy, associated to the presence of $3d^1$-Ti$^{3+}$ states., which is due to the presence of Ti ions with a Ti$^{3+}$ valence, a typical fingerprint of 2DEGs in titanates. Very strikingly, as shown in Fig. S8b, the intensity of the high energy satellite does not change going from normal (more bulk sensitive) to shallow emission (more surface sensitive), while the Ti$^{4+}$ features strongly decrease. This indicates that Ti$^{3+}$ valence states are at the surface. While these data are consistent with a TiO$_2$ termination, experimentally distinguishing the atomic surface termination needs further structural surface studies, like Grazing Incidence x-ray diffraction (GIXD), which should be done in future studies. It should be noted that the exact surface termination is also still an open question for the STO 2DEG [S4, S5].
In order to further support a TiO$_2$ termination of our ETO films, we show below band dispersion calculations for EuO terminated ETO. The calculations have been performed using exactly the same model and oxygen divacancy configuration described in the main text for TiO$_2$ terminated ETO (see inset of Fig. S8c).

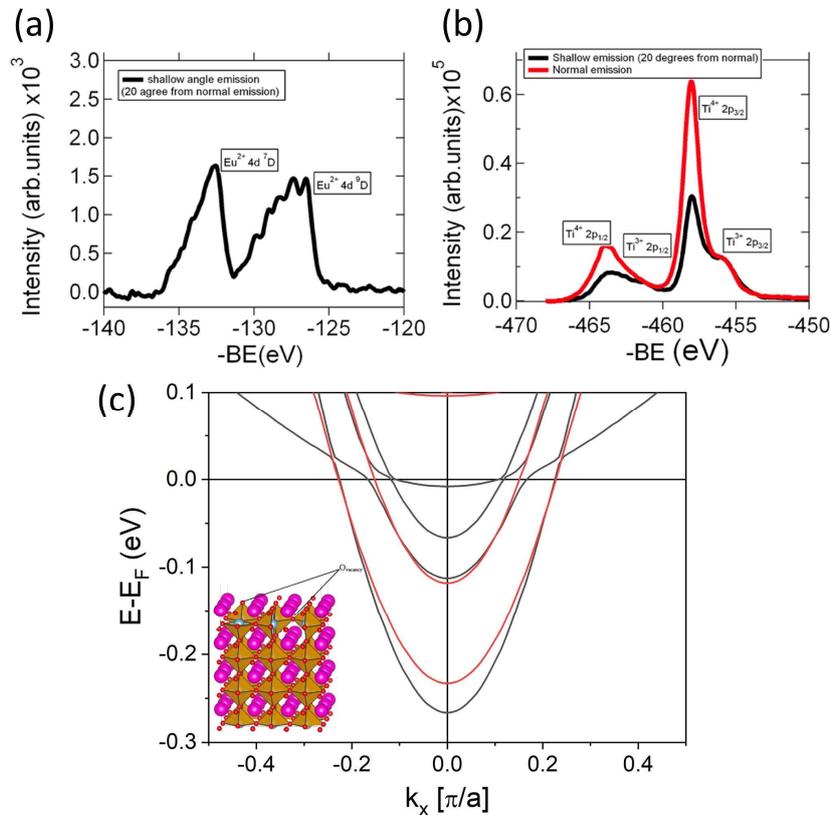

**Fig. S8:** Core level spectroscopy data of 15uc ETO deposited on buffer SrRuO$_3$: comparison between (a) Ti2p and (b) Eu4d core level spectra acquired in shallow emission angle (20 degrees from the surface normal) using incident photons at $h\nu$ = 600 eV. (c) Comparison between normal (red line) and shallow emission (black line) Ti2p core level spectra. (c) Calculated band dispersions around the Γ-point along Γ-X in EuO terminated 2x3x4 ETO (001) slab using PBESOL+U with same parameters employed for the TiO$_2$ termination and same oxygen divacancy.

One can see that also in the case the EuO termination the bands are spin-split, but the band bottom of the lowest lying band is of the order of 0.25 eV, about 1.5 times the one found experimentally. Moreover, heavy bands are above the Fermi level. Both results are in substantial disagreement with the experimental results, further confirming that the assumption of TiO$_2$ termination of our ETO films is possibly correct.

**S9- Spin-Orbit Coupling (SOC) in the in-plane di-vacancy configuration**

In Fig. S9 we report calculated band dispersions, with the SOC contribution, for ETO(001) and STO(001), considering configuration with an in-plane di-vacancy in a 2*a*x3*a*x4*a* supercell, described in the main text, with one missing oxygen in the topmost TiO$_2$ layer and a second one in the subsurface SrO/EuO layer at a distance of 6.2 Å of the STO(001) and ETO(001) surfaces, respectively (see Fig. 3a and 3b in the main text). See main text, Fig. 4a and 4b, for the analogous calculated band dispersions without SOC.

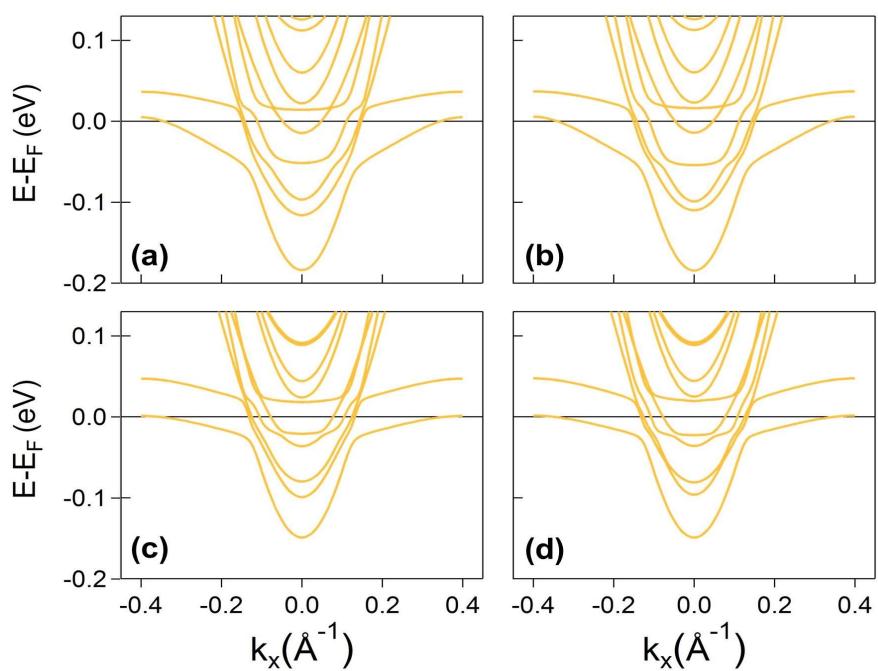

**Fig. S9**: Calculated band dispersions, including SOC, around the Γ-point along Γ-X in ETO (001) (upper panels) and STO (001) (bottom panels). a) and c) show calculations including SOC with magnetization direction along [001], while b) and d) show calculations including SOC with magnetization direction along [100]

**S10- Vertical di-vacancy configuration**

Among the different structural configurations, alternative to the one reported in the main text, on which we performed DFT+*U* calculation, we report here (Fig. S10) the results obtained considering a vertical divacancy in a 2*a*x2*a*x7*a* supercell, with missing oxygens in the first and second subsurface SrO/EuO layers of the STO(001) and ETO(001) surfaces, respectively.
Spin density integrated between -0.15 eV and the Fermi level, and layer- and element-resolved density of states, are also shown for both compounds. The calculated band dispersions (see comments in the main text) around the Γ-point along Γ-X and Γ-M directions are reported in the lower panels (Figs. S10g-l).

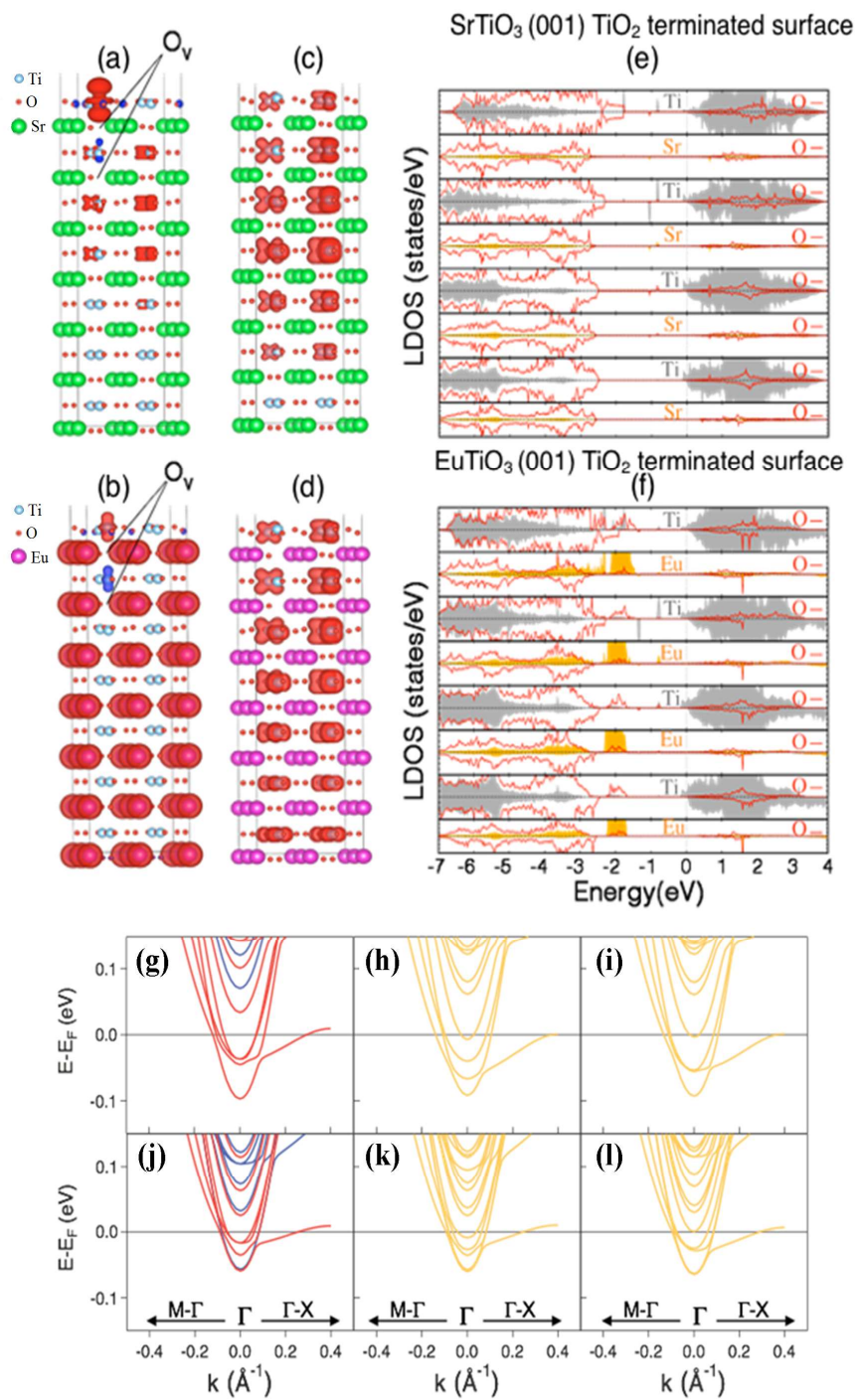

**Fig. S10**: side views of the relaxed (a) STO(001) and (b) ETO(001) surfaces with vertical oxygen divacancy in the first and second subsurface SrO/EuO layers. The panels show also the spin density with iso-values of 0.0005 e/Å$^3$ for the two systems. Red/blue corresponds to majority/minority spin density. (c)-(d) Spin density, integrated between -0.15 eV and the Fermi level of STO(001) (c) and ETO(001) (d). (e) and (f) Layer- and element-resolved density of states showing the spatial distribution and orbital polarization of the q2DEG in STO (001) and ETO(001), respectively. Calculated band dispersions around the Γ-point along Γ-X and Γ-M directions in ETO(001) (panels g,h,i) and STO(001) (panels j,k,l). (g) and (j) show the results without spin-orbit coupling (SOC) (red/blue correspond to majority/minority bands), h) and k) show calculations including SOC with magnetization direction along [001], while and l) show calculations including SOC with magnetization direction along [100].

**S11- Comparison between the DFT-U model planar di-vacancy configuration and experimental data**

In Fig. S11 we show a comparison between the ETO and STO (sample: STO-A) ARPES band dispersion data and DFT+U calculations (ground state) obtained considering configuration with an in-plane di-vacancy in a $2a\times3a\times4a$ supercell, considered in the main text, with one missing oxygen in the topmost $TiO_2$ layer and a second one in the subsurface SrO/EuO layer at a distance of 6.2 Å of the STO(001) and ETO(001) surfaces, respectively (see Fig. 4a and 4b in the main text).

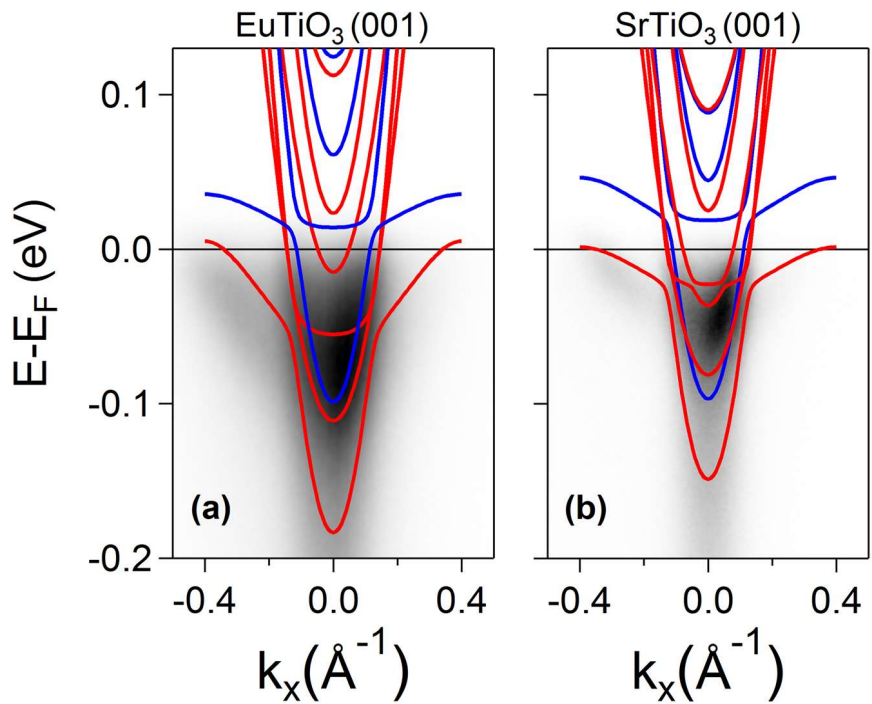

**Fig. S11**: Comparison between C+ ARPES data acquired at hν=85eV and DFT+$U$ calculations with the model described in the main text, around Γ-point along Γ-X direction, for (a) ETO(001) and (b) STO(001) (sample STO-A).

Supplementary References